# ADVERSARIAL-PLAYGROUND: A Visualization Suite Showing How Adversarial Examples Fool Deep Learning

Andrew P. Norton*    Yanjun Qi†

Department of Computer Science, University of Virginia

## ABSTRACT

Recent studies have shown that attackers can force deep learning models to misclassify so-called "adversarial examples:" maliciously generated images formed by making imperceptible modifications to pixel values. With growing interest in deep learning for security applications, it is important for security experts and users of machine learning to recognize how learning systems may be attacked. Due to the complex nature of deep learning, it is challenging to understand how deep models can be fooled by adversarial examples. Thus, we present a web-based visualization tool, ADVERSARIAL-PLAYGROUND, to demonstrate the efficacy of common adversarial methods against a convolutional neural network (CNN) system. ADVERSARIAL-PLAYGROUND is educational, modular and interactive. (1) It enables non-experts to compare examples visually and to understand why an adversarial example can fool a CNN-based image classifier. (2) It can help security experts explore more vulnerability of deep learning as a software module. (3) Building an interactive visualization is challenging in this domain due to the large feature space of image classification (generating adversarial examples is slow in general and visualizing images are costly). Through multiple novel design choices, our tool can provide fast and accurate responses to user requests. Empirically, we find that our client-server division strategy reduced the response time by an average of 1.5 seconds per sample. Our other innovation, a faster variant of JSMA evasion algorithm, empirically performed twice as fast as JSMA and yet maintains a comparable evasion rate[1].

**Index Terms:**   I.2.6 [Artificial Intelligence]: Learning—Connectionism and neural nets; K.6.5 [Management of Computing and Information Systems]: Security and Protection—Unauthorized access

## 1 INTRODUCTION

Investigating the behavior of machine learning systems in adversarial environments is an emerging topic at the junction of computer security and machine learning [1]. While machine learning models may appear to be effective for many security tasks like malware classification [7] and facial recognition [9], these classification techniques were not designed to withstand manipulations made by intelligent and adaptive adversaries. In contrast with applications of machine learning to other fields, security tasks involve adversaries that may respond maliciously to the classifier [1].

Recent studies show that intelligent attackers can force machine learning systems to misclassify samples by performing nearly imperceptible modifications to the sample before attempting classification [4, 10]. These samples, named as "adversarial examples," have effectively fooled many state-of-the-art deep learning models.

*e-mail: apn4za@virginia.edu
†e-mail: yanjun@virginia.edu

[1]Project source code and data from our experiments available at: https://github.com/QData/AdversarialDNN-Playground.

Adversarial examples for Deep Neural Network (DNN) models are usually crafted through an optimization procedure that searches for small, yet effective, perturbations of the original image (details in Sect. 2). Understanding why a DNN model performs as it does is quite challenging, and even moreso to understand how such a model can be fooled by adversarial examples.

With growing interest in adversarial deep learning, it is important for security experts and users of DNN systems to understand how DNN models may be attacked in face of an adversary. This paper introduces a visualization tool, ADVERSARIAL-PLAYGROUND, to enable better understanding of how different types of adversarial examples fool DNN systems. ADVERSARIAL-PLAYGROUND provides a simple and intuitive interface to let users visually explore the impact of three attack algorithms that generate adversarial examples. Users may specify parameters for a variety of attack types and generate new samples on-demand. The interface displays the resulting adversarial example compared to the original alongside classification likelihoods on both images from the DNN.

ADVERSARIAL-PLAYGROUND provides the following benefits:

- **Educational**: Fig. 1 shows a screen-shot of the visualization. This intuitive and simple visualization helps practitioners of deep learning to understand how their models misclassify and how adversarial examples by various algorithm differ.

- **Interactive**: We add two novel strategies in ADVERSARIAL-PLAYGROUND to make it respond to users' requests in a sufficiently quick manner. The interactive visualization allow users to gain deeper intuition about the behavior of DNN classification through maliciously generated inputs.

- **Modular**: Security experts can easily plug ADVERSARIAL-PLAYGROUND into their benchmarking frameworks as a module. The design also allows experts to easily add other DNN models or more algorithms of generating adversarial examples in the visualization.

To the authors' best knowledge, this is the first visualization platform showing how adversarial examples are generated and how they fool a DNN system.

The rest of this paper takes the following structure: Sect. 2 discusses the relevant backgrounds, Sect. 3 introduces the system organization and software design of ADVERSARIAL-PLAYGROUND, Sect. 4 presents an empirical evaluation with respect to different design choices, and Sect. 5 concludes the paper by discussing possible extensions.

## 2 BACKGROUND

### 2.1 Adversarial Examples and More

Studies regarding the behavior of machine learning models in adversarial environments generally fall into one of three categories: (1) *poisoning attacks*, in which specially crafted samples are injected into the training of a learning model, (2) *privacy-aware learning*, which aim to preserve the privacy of information in data samples, or (3) *evasion attacks*, in which the adversary aims to create inputs that are misclassified by a target classifier. Generating adversarial examples is part of this last category.

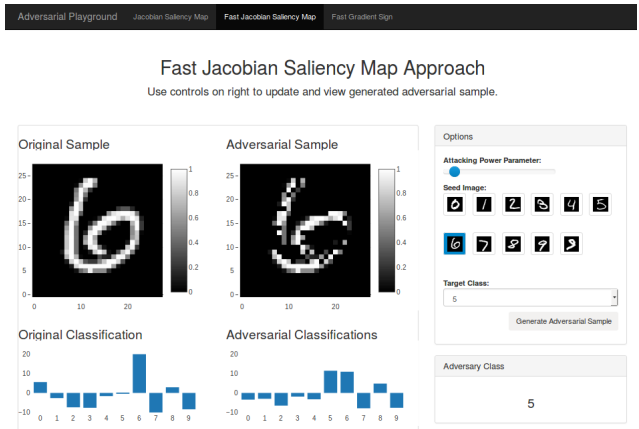

Figure 1: ADVERSARIAL-PLAYGROUND User Interface

The goal of adversarial example generation is to craft an input for a particular classifier that, while improperly classified, reveals only slight alteration on the input. To formalize the extent of allowed alterations, evasion algorithms minimize the difference between the "seed" input and the resulting adversarial example based on a predefined norm (a function measuring the distance between two inputs).

In some cases, the adversary specifies the "target" class of an adversarial sample — for example, the adversary may desire an image that looks like a "6" to be classified as a "5" (as in Fig. 1). This is referred to as a *targeted* attack. Conversely, if the adversary does not specify the desired class, the algorithm is considered to be *untargeted*.

Formally, let us denote $f : X \to C$ to be a classifier that maps the set of all possible inputs, $X$, to a finite set of classes, $C$. Then, given a target class $y_t \in C$, a seed sample $x \in X$, and a norm function $\|\cdot\|$, the goal of generating a targeted adversarial example is to find $x' \in X$ such that:

$$x' = \arg\min_{s \in X} \{\|x - s\| : f(s) = y_t\} \quad (1)$$

Similarly, in the untargeted case, the goal is to find $x'$ such that:

$$x' = \arg\min_{s \in X} \{\|x - s\| : f(s) \neq f(x)\} \quad (2)$$

In this formalization, we see there are two key degrees of freedom in creating a new evasion algorithm: targeted vs. untargeted attacks and the choice of norm functions. The latter category provides a useful grouping scheme for algorithms generating adversarial inputs, suggested by Carlini and Wagner [2]. ADVERSARIAL-PLAYGROUND uses two evasion algorithms provided by the `cleverhans` library [3]: the Fast Gradient Sign Method (FGSM) based on the $L^\infty$ norm, and the Jacobian Saliency Map Approach (JSMA) based on the $L^0$ norm [8].

### 2.2 DNNs and the MNIST Dataset

DNNs can efficiently learn highly-accurate models in many domains [5, 7]. Convolutional Neural Networks (CNNs), first popularized by LeCun et al. [6], perform exceptionally well on image classification. ADVERSARIAL-PLAYGROUND uses a state-of-the-art CNN model on the popular MNIST "handwritten digits" dataset for visualizing evasion attacks. This dataset contains 70,000 images of hand-written digits (0 through 9). Of these, 60,000 images are used as training data and the remaining 10,000 images are used for testing. Each sample is a $28 \times 28$ pixel, 8-bit grayscale image. Users of our system are presented with a collection of seed images, selected from each of the 10 classes in the testing set (see the right side of Fig. 1).

### 2.3 TensorFlow Playground

Our proposed package follows the spirit of *TensorFlow Playground* — a web-based educational tool that helps users understand how neural networks work [11]. TensorFlow Playground has been used in many classes as a pedagogical aid and helps the self-guided student learn more. Its impact inspires us to visualize adversarial examples through ADVERSARIAL-PLAYGROUND. Our web-based visualization tool assists users in understanding and comparing the impact of standard evasion techniques on deep learning models.

## 3 ADVERSARIAL-PLAYGROUND: A MODULAR AND INTERACTIVE VISUALIZATION SUITE

In creating our system, we made several design decisions to make ADVERSARIAL-PLAYGROUND educational, modular and interactive. Here, we present the four major system-level decisions we made: (1) building ADVERSARIAL-PLAYGROUND as a web-based application, (2) utilizing both client- and server-side code, (3) rendering images with the client rather than the server and (4) implementing a faster variation of JSMA attack.

We released all project code on GitHub in the interest of providing a high-quality, easy-to-use software package to demonstrate how adversarial examples fool deep learning.

### 3.1 A Web-based Visualization Interface

ADVERSARIAL-PLAYGROUND provides quick and effective visualizations of adversarial examples through an interactive webapp as shown by Fig. 1. The user selects one attacking algorithm from the navigation bar at the top of the webapp. On the right-hand pane, the user sets the attacking strength the algorithm using the slider, selects a seed image, and (if applicable) a target class. (Fig. 1 at right.) Selecting a seed image immediately loads the image to the left-hand display and displays the output of the CNN classifier in a bar chart below.

After setting the parameters, the user clicks "Generate Adversarial Sample." This runs the chosen adversarial algorithm in real-time to attempt generating an adversarial sample. The sample is displayed in the primary pane to the left of the controls (Fig. 1 at center). The generated sample is fed through the CNN classifier, and then the likelihoods are displayed in a bar chart below the sample. Finally, the classification of generated sample is displayed below the controls at right.

This web-based visualization generates adversarial examples "on-demand" from user-specified parameters. Therefore users can see the impact of different adversarial algorithms with varying configurations.

Developing ADVERSARIAL-PLAYGROUND as a web-based (as opposed to a local) application enables a large number of users to utilize the application without requiring an installation process on each computer. By eliminating an installation step, we encourage potential users who may be only casually interested in adversarial machine learning to explore what it is. This supports the pedagogical goals of the software package.

### 3.2 A Modular Design with Client-Server Division

Two key features of ADVERSARIAL-PLAYGROUND are its modular design and the division of the functionality between the client and server; the client handles user interaction and visualization, while the server handles more computationally intensive tasks. Fig. 2 diagrams the interaction between each component of our system. At the upper right, we have the user who may specify hyperparameters for the evasion algorithm. Moving counter-clockwise, these parameters are transferred to the server, where the appropriate adversarial algorithm module is selected and run against the pre-trained CNN module. TensorFlow is used to reduce computation time and improve compatibility. Finally, the resulting sample is sent to the client and plotted using the JavaScript library `Plotly.JS`.

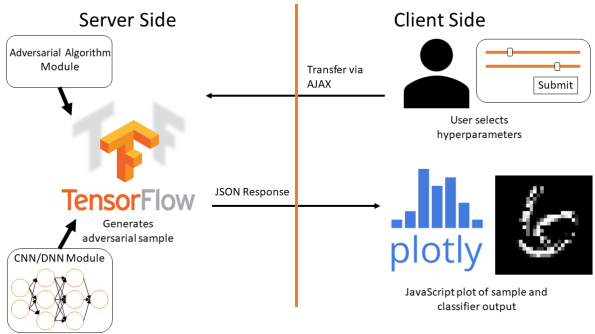

Figure 2: ADVERSARIAL-PLAYGROUND System Sketch

Users running a local copy of the webapp may easily customize the tool to their needs; by separating the deep learning model and the evasion methods from the main visualization and interface codebase, changing or adding DNN models or adding new adversarial algorithms is straightforward.

**TensorFlow based Server-side:** Our inspiration, the TensorFlow Playground, was written entirely in JavaScript and other client-side technologies, allowing a lightweight server to host the service for many users. Unfortunately, adversarial examples are usually generated on larger, deeper networks than those created by users of TensorFlow Playground, and this makes a JavaScript-only approach prohibitively slow.

Instead, we chose to use a GPU-enabled server running Python with TensorFlow to generate the adversarial examples on the back-end (server-side), then send the image data to the client. This provides increased speed (aiding interactivity), adds compatibility with other TensorFlow-based deep learning models and allows the flexibility of evasion algorithms (promoting modularity).

**Server-side Configuration:** The server-side of ADVERSARIAL-PLAYGROUND requires a computer with Python 3.5, TensorFlow 1.0 (or higher), the standard `SciPy` stack, and the Python package `Flask`. We have tested the code on Windows, Linux, and Mac operating systems.

To install, clone the GitHub repository and install the prerequisites via `pip3 -r install requirements.txt`. A pre-trained MNIST model is already stored in the GitHub repository; all that is needed to start the webapp is to run `python3 run.py`. Once the app is started, it will run on `localhost:9000`.

### 3.3 Visualizing Sample through Client-side Rendering

As shown in Fig. 2, through the client, the user adjusts hyperparameters and submits a request to generate an adversarial sample to the server. Once the TensorFlow back-end generates the adversarial image and classification likelihoods, the server returns this data to the client. Finally, this information is displayed graphically to the user through use of the `Plotly` JavaScript library.

As we generate adversarial samples on the server-side, it was tempting to produce the output images on the server as well. In our prototype, we used server-side rendering of these images with the Python library `matplotlib`, then downloaded the image for display on the client. However, we ultimately decided to assign all visualization tasks to the client, using JavaScript and the `Plotly.JS` library, after realizing this approach was faster.

This is because generating images on the server with the default `matplotlib` utilities required creating a full PNG image, writing it to disk, then transferring the image to the client; this took time and increased latency. Fortunately, client-side rendering of images required transmission of far less data; only pixel values for the $28 \times 28$ MNIST images and the 10 values for classification likelihoods needed to be sent. Additionally, the `Plotly.JS` library provided interactive plots that enable users to view the underlying values for each pixel. Empirically, switching to a client-side rendering of images reduced response time by approximately 1.5 seconds. (Sect. 4.2.)

ADVERSARIAL-PLAYGROUND's modularity extends into the visualization code, too. Although it may be hosted on any machine that supports TensorFlow, the web-based client/server division of the webapp allows the computationally intensive "back-end" to be hosted on a powerful server while the visualizations may be accessed from any device.

### 3.4 Faster Variant of JSMA Attack

While dividing the computation and visualization steps between the client and server saved some time, actually generating the adversarial example is where the most time is consumed (Table 1). In particular, the Jacobian Saliency Map Approach (JSMA) algorithm by Papernot et al. [8] can take more than two-thirds of a second to generate a single adversarial output. In order to provide an interactive experience, our web app must generate adversarial samples quickly. We therefore introduce a new, faster variant of the JSMA that maintains a comparable evasion rate to the original, but can take half as much time.

**JSMA Background:** Most state-of-the-art evasion algorithms are slow due to the expensive optimization and the large search space involved in image classification [2, 3].

The original JSMA algorithm is a targeted attack that uses the $L^0$ norm in Equation 1. To generate $x'$ from $x$, JSMA iteratively selects the "most influential" combination of two features to alter. To rank features by their influence, JSMA uses a *saliency map* of the forward derivative of the classifier. The ranking and alteration process is repeated until the altered sample is successfully classified as $y_t$ or the $L^0$ distance between the altered and seed samples exceeds a provided threshold, $\Upsilon$.

The largest consumption of time in JSAM is the combinatorial search over all feature pairs to determine the "best" pair to alter; if there are $M$ features in a given sample, JSMA must evaluate $\Theta(M^2)$ candidates at each iteration. When working on high-dimensional data, this can become prohibitively expensive. We introduce a new, faster variant of JSMA that maintains a comparable evasion rate to the original, which we call *Fast Jacobian Saliency Map Apriori* (FJSMA).

**FJSMA Improvement:** Our FJSMA approach is an approximation of JSMA that uses an *a priori* heuristic to significantly reduce the search space. Instead of considering all pairs of features $(p,q)$, our improvement only considers such pairs where $p$ is in the top $k$ features when ranked by the derivative in the $p$-coordinate, where $k$ is a small constant. (See red-bolded modifications to JSMA in Algorithm 1.)

If we denote the set consisting of the top $k$ elements in $A$ as ranked by $f$ by $\arg\text{top}_{x \in A}(f(x);k)$, then the loop in our Fast Jacobian Saliency Map Apriori (FJSMA) selection routine is $\Theta(k \cdot |\Gamma|)$, where $k \ll |\Gamma|$ and $|\Gamma| = M$ is the size of the feature set. Since determining the top $k$ features can be done in linear time, this is considerable improvement in asymptotic terms.

This modification improves the runtime from $\Theta(M^2)$ to $\Theta(M \cdot k)$, where $M$ is the feature size and $k$ is some small constant. Our experiments show $k$ may be as little as 15% of $M$ and still maintain the same efficacy in terms of evasion rate as JSMA.

## 4 PERFORMANCE TESTING

We conducted a series of timing tests to quantify how our design choices have influenced the speed of interactive responses by ADVERSARIAL-PLAYGROUND. First, we consider the impact of relegating the visualization code to the client (from Sect. 3.2); then,

**Algorithm 1** Fast Jacobian Saliency Map Apriori Selection
$\nabla \mathbf{F}(\mathbf{X})$ is the forward derivative, $\Gamma$ the features still in the search space, $t$ the target class, and $k$ is a small constant

**Input:** $\nabla \mathbf{F}(\mathbf{X}), \Gamma, t, k$

1: $K = \arg\text{top}_{p \in \Gamma} \left( -\frac{\partial \mathbf{F}_t(\mathbf{X})}{\partial \mathbf{X}_p}; k \right)$  ▷ Changed for FJSMA
2: **for** each pair $(p, q) \in K \times \Gamma, p \neq q$ **do**  ▷ Changed for FJSMA
3:   $\alpha = \sum_{i=p,q} \frac{\partial \mathbf{F}_t(\mathbf{X})}{\partial \mathbf{X}_i}$
4:   $\beta = \sum_{i=p,q} \sum_{j \neq t} \frac{\partial \mathbf{F}_j(\mathbf{X})}{\partial \mathbf{X}_i}$
5:   **if** $\alpha < 0$ and $\beta > 0$ and $-\alpha \times \beta > \max$ **then**
6:     $p_1, p_2 \leftarrow p, q$
7:     $\max \leftarrow -\alpha \times \beta$
8:   **end if**
9: **end for**
10: **return** $p_1, p_2$

we show that FJSMA is faster and just as accurate as the JSMA implementation by `cleverhans` package.

### 4.1 Client-side Rendering Improves Response Speed

| Rendering done on... | Server | Image Download | Total |
|---|---|---|---|
| Server-side | 4472 | 350 | 4821 |
| Client-side | 3335 | — | 3335 |
| **Difference** | 1137 | 350 | 1486 |

Table 1: Latency with and without client-side visualization. A time profiling of the latency experienced by the user when 1) the server handled all computation and visualization and 2) the visualization was offloaded to the client. The "Server" column denotes time taken for the server to respond, while the "Image Download" column shows the additional time taken to transfer each image (only applicable for server-side rendering).

We first conducted timing tests to evaluate how the choice of client-side rendering (Sect. 3.2) has influenced the speed of responding to users' requests. We loaded the webapp and measured the response time of the server for a variety of seed images and target classes. We repeated these for a total of between 10 and 16 times (depending on the algorithm), averaged the response time, and reported the result in Table 1. The majority of the time for both with and without client-side visualization is in the server computation. However, offloading the visualization to the client resulted in a nearly 1.5-second speedup (an approximately 30% difference). Interestingly, not only did the image download time get eliminated, but the server computation time was reduced as well. This is, in part, due to the reduction of I/O operations and image generation required by the server when visualization is done by the client.

### 4.2 FJSMA Improvement

We propose a faster approximation of the JSMA attacking algorithm: FJSMA in Sect. 3.4. Using the same CNN model we used in ADVERSARIAL-PLAYGROUND for the MNIST dataset, we compared FJSMA with JSMA through two metrics: (1) the "wall clock" time needed for successfully generating an adversarial example, and (2) the *evasion rate* — a standard metric that reports the percentage of seed images that were successfully converted into adversarial samples.

This comparison was conducted in a batch manner. We ran both evasion attacks on the 10000-sample MNIST testing set for a range of values of the $\Upsilon$ parameter for both algorithms. For FJSMA, we also varied the value of input parameter $k$ (the percentage of the feature-set size). Intuitively, this $k$ value is a control on how tight of an approximation FJSMA is to JSMA; as $k$ grows larger, we should expect the performance of the two approaches to converge to each other.

| $\Upsilon$ | 10% | 15% | 20% | 25% |
|---|---|---|---|---|
| JSMA Evasion Rate | 0.658 | 0.824 | 0.867 | 0.879 |
| FJSMA Evasion Rate [$k = 10\%$] | 0.583 | 0.777 | 0.823 | 0.826 |
| FJSMA Evasion Rate [$k = 15\%$] | 0.613 | 0.816 | 0.867 | 0.871 |
| FJSMA Evasion Rate [$k = 20\%$] | 0.633 | 0.833 | 0.878 | 0.887 |
| FJSMA Evasion Rate [$k = 30\%$] | 0.638 | 0.844 | 0.896 | 0.901 |
| JSMA Time (s) | 0.606 | 0.745 | 0.807 | 0.803 |
| FJSMA Time [$k = 10\%$] (s) | 0.411 | 0.468 | 0.490 | 0.485 |
| FJSMA Time [$k = 15\%$] (s) | 0.414 | 0.473 | 0.483 | 0.484 |
| FJSMA Time [$k = 20\%$] (s) | 0.415 | 0.466 | 0.482 | 0.483 |
| FJSMA Time [$k = 30\%$] (s) | 0.415 | 0.464 | 0.490 | 0.485 |

Table 2: JSMA and FJSMA Comparison. Each column represents a test run with a particular value of $\Upsilon$ (the maximum allowed perturbation for a sample). The top half of the table provides the average evasion rate for each algorithm, while the bottom half provides the average time (in seconds) that it took to generate an adversarial example. The FJSMA algorithm was run with multiple values of $k$, where $k$ was $10\%, 15\%, 20\%,$ and $30\%$ of the feature space size.

Results of this experiment are summarized in Table 2. The `cleverhans` JSMA and the proposed FJSMA attack achieve similar evasion rates for all tested values of $\Upsilon$ and $k$, with larger values of $k$ increasing the evasion rate. Curiously, for $k \geq 20\%$, our implementation of FJSMA even outperforms that of `cleverhans` JSMA; this is likely due to implementation details. The average time to form an evasive sample from a seed benign sample is given in the second half of the table. Our FJSMA approach greatly improves upon the speed of JSMA. However, varying the value of $k$ does not produce a significant variation in runtime per sample; we conjecture this is because of the small feature space of MNIST and that searching 30% of the feature space likely does not dominate the runtime. In summary, FJSMA achieves a significant improvement in speed, while maintaining essentially the same evasion rate — an important advantage for interactive visualization.

## 5 DISCUSSION AND FUTURE WORK

The study of evasion attacks on machine learning models is a rapidly growing field. In this paper, we present a web-based tool ADVERSARIAL-PLAYGROUND for visualizing the performance of adversarial examples against deep neural networks. ADVERSARIAL-PLAYGROUND enables non-experts to compare adversarial examples visually and can help security experts explore more vulnerability of deep learning. It is modular and interactive. To our knowledge, our platform is the first visualization-focused package for adversarial machine learning.

A straightforward extension of this work is to increase the variety of supported evasion methods. For example, including the new attacks based on $L^0$, $L^2$, and $L^\infty$ norms from Carlini and Wagner's recent paper [2] would be a good next step in comparing the performance of multiple evasion strategies. However, expansion in this manner presents an additional issue of latency. To generate evading samples "on-demand," the adversarial algorithm must run quickly; these other algorithms take much longer to execute than those we selected, so some time-saving techniques must be explored.

Another direction for development is to provide more choices of classifiers and datasets. Allowing the user to select from CIFAR, ImageNet, and MNIST data would highlight the similarities and differences between how a single attack method deals with different data. Similarly, providing the user with a choice of multiple pre-trained models — possibly hardened against attack through adversarial training — would help distinguish artifacts of model choice from the behavior of the attack. These two extensions would help users more fully understand the behavior of an adversarial algorithm.

## REFERENCES

[1] M. Barreno, B. Nelson, A. D. Joseph, and J. Tygar. The Security of Machine Learning. *Machine Learning*, 81(2):121–148, 2010.

[2] N. Carlini and D. Wagner. Towards evaluating the robustness of neural networks. *CoRR*, abs/1608.04644, 2016.

[3] I. J. Goodfellow, N. Papernot, and P. D. McDaniel. cleverhans v0.1: an adversarial machine learning library. *CoRR*, abs/1610.00768, 2016.

[4] I. J. Goodfellow, J. Shlens, and C. Szegedy. Explaining and harnessing adversarial examples. *arXiv preprint arXiv:1412.6572*, 2014.

[5] A. Krizhevsky, I. Sutskever, and G. E. Hinton. ImageNet Classification with Deep Convolutional Neural Networks. In *Advances in Neural Information Processing Systems*, pp. 1097–1105, 2012.

[6] Y. LeCun, L. Bottou, Y. Bengio, and P. Haffner. Gradient-based learning applied to document recognition. *Proceedings of the IEEE*, 86(11):2278–2324, November 1998.

[7] Microsoft Corporation. Microsoft Malware Competition Challenge. https://www.kaggle.com/c/malware-classification, 2015.

[8] N. Papernot, P. D. McDaniel, S. Jha, M. Fredrikson, Z. B. Celik, and A. Swami. The limitations of deep learning in adversarial settings. *CoRR*, abs/1511.07528, 2015.

[9] O. M. Parkhi, A. Vedaldi, and A. Zisserman. Deep Face Recognition. In *British Machine Vision Conference*, 2015.

[10] C. Szegedy, W. Zaremba, I. Sutskever, J. Bruna, D. Erhan, I. J. Goodfellow, and R. Fergus. Intriguing properties of neural networks. *CoRR*, abs/1312.6199, 2013.

[11] J. Yosinski, J. Clune, A. M. Nguyen, T. J. Fuchs, and H. Lipson. Understanding neural networks through deep visualization. *CoRR*, abs/1506.06579, 2015.